\documentclass[preprint]{ptephy}

\preprintnumber{HUPD-1402} 

\usepackage{natbib}
\def\slash#1{\not\!#1}
\begin{document}

\title{Gauge covariant solution for the Schwinger-Dyson equation in three-dimensional 
QED with Chern-Simons term}

\author{\name{Yuichi Hoshino}{1\dag}, \name{Tomohiro Inagaki}{2,\ast, \dag}, and \name{Yuichi Mizutani}{2,}\thanks{These authors contributed equally to this work}}

\address{\affil{1}{Kushiro National College of Technology, Otanoshike-Nishi 2-32-1, Kushiro 084-0916, Japan}
\affil{2}{Information Media Center, Hiroshima University, Higashi-Hiroshima 739-8521, Japan}
\email{inagaki@hiroshima-u.ac.jp}}

\begin{abstract}%
An Abelian gauge theory with Chern-Simons (CS) term is investigated for a four-component Dirac fermion in 1+2 dimensions. The Ball-Chiu (BC) vertex function is employed to modify the rainbow-ladder approximation for the Schwinger-Dyson (SD) equation. We numerically solve the SD equation and show the gauge dependence for the resulting phase boundary for the parity and the chiral symmetry.
\end{abstract}

\subjectindex{B02, B04}

\maketitle


\section{Introduction}

Gauge theories provide a well-defined description for interactions between elementary particles. The theories can be applied to a wide variety of systems. Lots of works have been done to investigate the phase structure of gauge theories. The phase structure can be found by evaluating the expectation value of the order parameters under the ground state. Non-perturbative aspects of the gauge theories are an essential piece to induce a non-trivial phase structure. In this paper we employ the SD equation as a non-perturbative approach to calculate the expectation value of the order parameters. 

The SD equation gives general relationships between full Green functions. The full fermion propagator can be described by the full gauge boson propagator and the full vertex functions between the fermion and the gauge field. The SD equation is formed as a set of infinite coupled equations. In the rainbow-ladder approximation the full gauge boson propagator and the full vertex functions are replaced by the tree level ones. The approximation provides a closed form expression for the full fermion propagator. It also introduces a redundant gauge dependence of the solution. For a numerical analysis the Landau gauge is often used, since it simplifies the expression and preserves the Ward identity, $Z_1=Z_2$, for Abelian gauge theory in 1+3 dimensions at zero temperature.

The rainbow-ladder approximation is also applied to the SD equation for 1+2 dimensional gauge theories. A finite CS term violates the Ward identity under a constant gauge parameter in the rainbow-ladder approximation. The validity of the approximation is not clear to take the Landau gauge. Alternative approaches have been investigated to find the phase structure of 1+2 dimensional gauge theories with CS term. The phase structure has been evaluated by taking the non-local gauge \cite{Ebihara:1994wm, Kondo:1994bt, Kondo:1994cz} and find the first order phase boundary.

The origin of the gauge dependence is introduced by the rainbow-ladder approximation. The Ward identity can be preserved by taking a general form for the vertex function \cite{Ball:1980ay}. It is expected to reduce the gauge dependence for the solution of the approximated SD equation. The BC vertex form is used to evaluate the SD equation in 1+3 dimensional QED \cite{Curtis:1990zs, Bashir:1994az}. It is also adopted to 1+2 dimensional QED without CS term \cite{Fischer:2004nq, Bashir:2005wt, Bashir:2011vg,Zhu:2013zna}.

One of the interesting phenomena is found for low dimensional fermions with a gauge interaction. In 1+2 dimensions a topological CS term is naturally introduced. The term generates a topological mass for the gauge field and realizes a complex phase structure. In this paper we consider the four components Dirac fermions in QED with CS term. The chiral and parity invariances are assumed for the theory and the phase structure of the theory is investigated. In Sec. 2 we briefly review the SD equation and BC vertex form in 1+2 dimensional QED with CS term. In Sec. 3 the SD equation is numerically solved with BC vertex. We show the gauge dependence in the solution. The result is compared with the one obtained under the rainbow-ladder approximation. In Sec.4 we give some concluding remarks.

\section{Schwinger-Dyson equation and BC vertex form}

Throughout this paper we consider an Abelian gauge theory, especially quantum electrodynamics (QED), with massless fermion. 
In 1+2 dimension a two-component spinor gives an irreducible representation for a fermion field.
It is enough to represent two spin states for the fermion. 
However, it is also interesting to consider a four-component spinor which is constructed by two kinds of two-component spinors.
A chiral symmetry can be defined for the four-component spinor even in 1+2 dimension.

A general form of Lagrangian for a four-component Dirac fermion with QED interaction is given by
\begin{eqnarray}
{\cal L}&=&-\frac{1}{4}F_{\mu\nu}F^{\mu\nu}-\frac{1}{2\alpha}(\partial_{\mu}A^{\mu})^2
+\frac{\mu}{2}\varepsilon^{\mu\nu\rho} A_{\mu} \partial_{\nu} A_{\rho}
\nonumber \\
&&+\bar{\psi}(i \!\!\not\partial - e\gamma^{\mu}A_{\mu} - m \tau)\psi ,
\label{Lagrangian}
\end{eqnarray}
where $F_{\mu\nu}$ is a field strength,
$F_{\mu\nu}\equiv\partial_{\mu}A_{\nu}-\partial_{\nu}A_{\mu}$, 
$\alpha$ represents the gauge parameter in Lorentz gauge.
The third term in the first line, CS term, violates the invariance under the parity and the time reversal transformation. The term induces a topological mass, $\mu$, for the gauge field, $A_{\mu}$. 
In 1+2 dimensions the mass dimensions for the gauge field and the fermion field are $1/2$ and $1$, respectively. 
Thus the gauge coupling, $e$, has a unity mass dimension.

The field, $\psi$, describes a massless four-component Dirac fermion and $\gamma^{\mu} (\mu \in \{0,1,2\})$ represents $4\times 4$ matrices which satisfy the Clifford algebra $\{\gamma^\mu,\gamma^\nu\}=2g^{\mu\nu}$. 
There are two Casimir operators which anti-commute with all $\gamma^{\mu} (\mu \in \{0,1,2\})$. We write the operators as $\gamma^3$, $\gamma^5$ and define $\tau\equiv -i\gamma^3\gamma^5$ \cite{Ebihara:1994wm, Bur:1992}.
An explicit expression for these matrices is given by
\begin{eqnarray}
&& \gamma^0=\left(
\begin{array}{cc}
\sigma^3 & 0 \\
0 & -\sigma^3
\end{array}
\right), \, 
\gamma^1=-i \left(
\begin{array}{cc}
\sigma^1 & 0 \\
0 & -\sigma^1
\end{array}
\right), \, 
\gamma^2=-i\left(
\begin{array}{cc}
\sigma^2 & 0 \\
0 & -\sigma^2
\end{array}
\right), \, 
\nonumber \\
&& \gamma^3=\left(
\begin{array}{cc}
0 & I \\
I & 0
\end{array}
\right), \, 
\gamma^5=-i \left(
\begin{array}{cc}
0 & I \\
-I & 0
\end{array}
\right), \, 
\tau=-i\gamma^3\gamma^5=\left(
\begin{array}{cc}
I & 0 \\
0 & -I
\end{array}
\right),
\end{eqnarray} 
where $\sigma^{i} (i=1,2,3)$ denotes the Pauli matrices.

The Lagrangian is invariant under two types of global chiral transformations,
\begin{eqnarray}
\psi & \rightarrow & e^{i \theta_3 \gamma^3}\psi,
\nonumber \\
\psi & \rightarrow & e^{i \theta_5 \gamma^5}\psi,
\label{transformation:chiral}
\end{eqnarray}
where $\theta_3$ and $\theta_5$ are constant parameters.
Above chiral symmetries prevent the Lagrangian from having mass terms proportional to $\bar{\psi}\psi$, $\bar{\psi}\gamma^3\psi$ and $\bar{\psi}\gamma^5\psi$. The chiral symmetries also prohibit kinetic terms proportional to $\bar{\psi}\partial_\mu\gamma^\mu\gamma^3\psi$, $\bar{\psi}\partial_\mu\gamma^\mu\gamma^5\psi$ and $\bar{\psi}\partial_\mu\gamma^\mu\gamma^3\gamma^5\psi$.

The parity transformation is defined by the flip in the sign of the spatial coordinate, $(x,y) \rightarrow (-x,y)$.
The parity transformation for the fermion field is given by 
\begin{eqnarray}
\psi(x,y) & \rightarrow & i \gamma_1 \gamma_3 e^{i \theta_p \tau}\psi(-x,y) .
\label{transformation:parity}
\end{eqnarray}
Because of the CS term, the Lagrangian (\ref{Lagrangian}) is not invariant under the parity transformation.
We can define an parity breaking mass term proportional to $\bar{\psi}\tau\psi$ in 1+2 dimensions. The term is invariant under the chiral transformations (\ref{transformation:chiral}). 

We consider that the 1+2 dimensional QED with the CS term is a good place to study the SD analysis of higher dimensional gauge theories. It is also interesting in a context of condensed matter physics.
The CS gauge theory is applied to describe a fractional quantum Hall effect. Anyons can be arise from the CS term.
The chiral symmetry can be observed only for a massless fermion. 
In an electron system the massless fermion is found in graphene. 
We means that the the energy momentum dispersion satisfies a linear relation. 
If the fermion has non-vanishing mass, 
it modifies the dispersion relation and the chiral symmetry is broken.
The parity can be defined for a four-component fermion. It has been introduced in condensed matter physics in Ref.\cite{Sem:1984}.
From the two degeneracy points per Brillouin zone two-component fermion, $\psi_1$, has its doubler, $\psi_2$. 
The parity is defined as a transformation between $\psi_1$ and $\psi_2$. It can be rewritten by a single four-component
fermion.
It has been applied for a massless fermion system on graphene in Ref.\cite{Her:2009}.
The graphenefs honeycomb lattice is defined with a four-component fermion. 
The Lagrangian has a global SU(2) chiral symmetry generated by $\{\gamma_3, \gamma_5, \tau\}$.
If both chirality and parity breaking fermion masses proportional to $\bar{\psi}\psi$ and $\bar{\psi}\tau\psi$
develop non-vanishing values, the SU(2) symmetry breaks down to the U(1) symmetry generated by $\tau$. 

To study the phase structure of the theory we employ the SD equation.
The SD equation for the fermion self-energy, $\Sigma(p)$, reads
\begin{eqnarray}
-i\Sigma(p) = (-i e)^2 \int \frac{d^3 k}{(2\pi)^3} 
\gamma^{\mu} D_{\mu\nu}(p-k) S(k) \Gamma^{\nu}(p,k),
\label{SD}
\end{eqnarray}
where $D_{\mu\nu}(p-k)$ and $S(k)$ are the gauge boson and the fermion
propagators, respectively. $\Gamma^{\nu}(p,k)$ represents the vertex function between the fermion and the gauge boson. 
The fermion self-energy is defined through
\begin{eqnarray}
S(p) = \frac{i}{\slash{p}-m\tau-\Sigma(p)+i\varepsilon}
=\frac{i}{A(p)\slash{p}  - B(p)}.
\end{eqnarray}
In this equation $A(p)$ and $B(p)$ have $4\times 4$ bi-spinor forms.
It is more convenient to decompose the fermion propagator,
\begin{eqnarray}
S(p)=\frac{i}{A_{+}(p)\slash{p}  - B_{+}(p) }\chi_{+}
 + \frac{i}{A_{-}(p)\slash{p}  - B_{-}(p)}\chi_{-},
\label{pro:fermion}
\end{eqnarray}
where $\chi_{\pm}$ are the chirality projection operators
\begin{eqnarray}
\chi_{\pm}\equiv \frac{1 \pm \tau}{2}.
\end{eqnarray}

The gauge boson propagator, $D_{\mu\nu}(p-k)$, and the vertex function, $\Gamma^{\nu}(p,k)$, are necessary to solve Eq.~(\ref{SD}). In the rainbow-ladder approximation $D_{\mu\nu}(p-k)$ and $\Gamma^{\nu}(p,k)$ are replaced by the tree level ones. Here we fix the gauge boson propagator to the tree level one,
\begin{eqnarray}
D_{\mu\nu}(p) &=& \frac{-i}{p^2-\mu^2} 
\left(  g_{\mu\nu} - \frac{p_{\mu} p_{\nu} }{p^2} \right) 
- \mu \frac{1}{p^2 - \mu^2} \frac{1}{p^2}\varepsilon_{\mu\nu\rho} p^{\rho}
-i \alpha \frac{p_{\mu} p_{\nu} }{p^4},
\end{eqnarray}
The CS term induces a topological mass for the gauge boson. 

The tree level vertex function, $\Gamma^{\nu}(p,k)=\gamma^{\nu}$ does not keep the Ward-Takahashi (WT) identity,
\begin{eqnarray}
i(p-k)_{\mu}{\it \Gamma}^{\mu}(p,k) = S^{-1}(k) - S^{-1}(p),
\label{WT}
\end{eqnarray}
for a constant gauge parameter, $\alpha$. We wish to keep the identity to find a gauge covariant analysis.

The vertex function can be decomposed into the transverse and the longitudinal parts,
\begin{eqnarray}
\Gamma^\mu = \Gamma_{T}^{\mu} + \Gamma_{L}^{\mu}.
\end{eqnarray}
Substituting Eq. (\ref{pro:fermion}) to Eq. (\ref{WT}), we find that the WT identity is satisfied by the BC vertex function \cite{Ball:1980ay},
\begin{eqnarray}
\Gamma_{L}^{\mu} = \Gamma_{L+}^{BC \mu} + \Gamma_{L-}^{BC \mu},
\end{eqnarray}
where
\begin{eqnarray}
\Gamma_{L\pm}^{BC \mu} &=& \biggl\{ \frac{1}{2} (A_{\pm}(p) + A_{\pm}(k))\gamma^{\mu}
+ \frac{1}{2}(A_{\pm}(p) - A_{\pm}(k)) \frac{p^{\mu} + k^{\mu}}{p^2 - k^2}
(\slash{p} + \slash{k}) 
\nonumber\\
&-& (B_{\pm}(p) - B_{\pm}(k)) \frac{p^{\mu} + k^{\mu}}{p^2 - k^2} \biggr\} \chi_{\pm}.
\label{BC}
\end{eqnarray}
Since the transverse part of the vertex function satisfies,
\begin{eqnarray}
i(p-k)_{\mu}{\it \Gamma}_{T}^{\mu}(p,k) =0,
\end{eqnarray}
no constraint is derived from the Ward-Takahashi identity for $\Gamma_{T}^{\mu}$. Below we employ a simple form,
\begin{eqnarray}
\Gamma^{\mu}=\Gamma_{L+}^{BC \mu}+\Gamma_{L-}^{BC \mu},\,\, \Gamma_{T}^{\mu} = 0.
\end{eqnarray}

We substitute the BC vertex function (\ref{BC}) in Eq.~(\ref{SD}).  Performing the Wick rotation and integrating the angle variables, we obtain the following simultaneous equations:
\begin{eqnarray}
A_{\pm}(p) &=& 1- \frac{e^2}{8\pi^2} \int dk \frac{k}{A_{\pm}^2(k) k^2 + B_{\pm}^2(k)}
\nonumber\\
&&\times \biggl\{ F_{A\pm}(k,p,\mu) 
 + G_{A1\pm}(k,p,\mu) \ln \! \frac{(p-k)^2}{(p+k)^2} 
 + G_{A2\pm}(k,p,\mu) \ln \! \frac{(p-k)^2 + \mu^2}{(p+k)^2 + \mu^2} \nonumber\\
&& + G_{A3\pm}(k,p,\mu) 
\biggl( \ln \! \frac{(p-k)^2}{(p+k)^2} - \ln \! \frac{(p-k)^2 + \mu^2}{(p+k)^2 + \mu^2} \biggr)
\biggr\},
\label{SD:A:BC}
\end{eqnarray}
with
\begin{eqnarray}
F_{A\pm}(k,p,\mu) &=& (1+\alpha) \frac{k}{p^2} A_{\pm}(k)(A_{\pm}(p) + A_{\pm}(k))\nonumber\\
&&+\frac{k}{p^2 (p^2 - k^2)} \bigl\{
 - (1 - \alpha ) (p^2+k^2) A_{\pm}(k)(A_{\pm}(p) - A_{\pm}(k)) \nonumber\\
&& \pm \mu B_{\pm}(k) (A_{\pm}(p) - A_{\pm}(k)) 
 \mp 2 \mu A_{\pm}(k) (B_{\pm}(p) - B_{\pm}(k)) \nonumber\\
&& - 2(1 - \alpha) B_{\pm}(k) (B_{\pm}(p) - B_{\pm}(k) )
\bigr\},
\end{eqnarray}
\begin{eqnarray}
G_{A1\pm}(k,p,\mu) &=& \frac{\alpha}{2p^3} \biggl\{
 \frac{p^2+k^2}{2} A_{\pm}(k)(A_{\pm}(p) + A_{\pm}(k)) \nonumber\\
&& + \frac{p^2 - k^2}{2} A_{\pm}(k) (A_{\pm}(p) - A_{\pm}(k)) 
 - B_{\pm}(k) ( B_{\pm}(p) - B_{\pm}(k) )
\bigr\},
\end{eqnarray}
\begin{eqnarray}
G_{A2\pm}(k,p,\mu) &=& \frac{\mu}{2p^3} \biggl\{
 \frac{\mu}{2} A_{\pm}(k)(A_{\pm}(p) + A_{\pm}(k)) 
  \pm B_{\pm}(k) (A_{\pm}(p) + A_{\pm}(k)) \biggr\} \nonumber\\
&& +\frac{\mu^2 + 2(p^2+k^2)}{2 p^3 (p^2-k^2)} \biggl\{
 - \frac{ p^2 + k^2 }{2} A_{\pm}(k)(A_{\pm}(p) - A_{\pm}(k)) \nonumber\\
&& \pm \frac{\mu}{2} B_{\pm}(k) (A_{\pm}(p) - A_{\pm}(k)) 
 \mp \mu A_{\pm}(k) (B_{\pm}(p) - B_{\pm}(k) ) \nonumber\\
&&- B_{\pm}(k) (B_{\pm}(p) - B_{\pm}(k) ) \biggr\},
\end{eqnarray}
\begin{eqnarray}
G_{A3\pm}(k,p,\mu) &=& \frac{p^2-k^2}{2\mu^{2}p^{3}} \biggl\{
 \frac{p^2-k^2}{2} A_{\pm}(k)(A_{\pm}(p) + A_{\pm}(k)) \nonumber\\
&& + \frac{ p^2 +k^2 }{2} A_{\pm}(k) (A_{\pm}(p) - A_{\pm}(k)) \nonumber\\
&& \pm \mu B_{\pm}(k) (A_{\pm}(p) + A_{\pm}(k)) 
 \pm \frac{\mu}{2} B_{\pm}(k) (A_{\pm}(p) - A_{\pm}(k)) \nonumber\\
&&+ B_{\pm}(k) (B_{\pm}(p) - B_{\pm}(k) ) \biggr\}.
\end{eqnarray}
and
\begin{eqnarray}
B_{\pm}(p) &=& \pm m - \frac{e^2}{8\pi^2} \int dk \frac{k}{A_{\pm}^2(k) k^2 + B_{\pm}^2(k)}
\nonumber\\
&&\times \biggl\{ F_{B\pm}(k,p,\mu) 
 + G_{B1\pm}(k,p,\mu) \ln \! \frac{(p-k)^2}{(p+k)^2} 
  + G_{B2\pm}(k,p,\mu) \ln \! \frac{(p-k)^2 + \mu^2}{(p+k)^2 + \mu^2} \nonumber\\
&& + G_{B3\pm}(k,p,\mu) 
\biggl( \ln \! \frac{(p-k)^2}{(p+k)^2} 
       - \ln \! \frac{(p-k)^2 + \mu^2}{(p+k)^2 + \mu^2} \biggr)
\biggr\},
\label{SD:B:BC}
\end{eqnarray}
with
\begin{eqnarray}
F_{B\pm}(k,p,\mu) &=& \frac{k}{p^2 - k^2} \bigl\{
\pm \mu A_{\pm}(k)(A_{\pm}(p) - A_{\pm}(k)) \nonumber\\
&& + 2(1-\alpha) B_{\pm}(k) (A_{\pm}(p) - A_{\pm}(k)) \nonumber\\
&& - 2(1-\alpha) A_{\pm}(k) (B_{\pm}(p) - B_{\pm}(k)) \bigr\}, 
\end{eqnarray}
\begin{eqnarray}
G_{B1\pm}(k,p,\mu) &=& \frac{\alpha}{2p} \bigl\{
 B_{\pm}(k)(A_{\pm}(p) + A_{\pm}(k))  + A_{\pm}(k) (B_{\pm}(p) - B_{\pm}(k)) \bigr\},
\end{eqnarray}
\begin{eqnarray}
G_{B2\pm}(k,p,\mu) &=& \frac{1}{p} \biggl\{
 \pm \frac{\mu}{2} A_{\pm}(k)(A_{\pm}(p) + A_{\pm}(k)) 
  + B_{\pm}(k) (A_{\pm}(p) + A_{\pm}(k)) \biggr\}\nonumber\\
&+& \frac{\mu^2 + 2(p^2+k^2)}{2p(p^2-k^2)} \biggl\{
 \pm \frac{\mu}{2} A_{\pm}(k)(A_{\pm}(p) - A_{\pm}(k)) \nonumber\\
&& + B_{\pm}(k) (A_{\pm}(p) - A_{\pm}(k)) 
- A_{\pm}(k) (B_{\pm}(p) - B_{\pm}(k) ) \biggr\},
\end{eqnarray}
\begin{eqnarray}
G_{B3\pm}(k,p,\mu) &=& -\frac{p^2-k^2}{2\mu^2p} \bigl\{
 \pm \mu A_{\pm}(k)(A_{\pm}(p) + A_{\pm}(k)) 
  \pm \frac{\mu}{2} A_{\pm}(k) (A_{\pm}(p) - A_{\pm}(k)) \nonumber\\
&&+ B_{\pm}(k) (A_{\pm}(p) - A_{\pm}(k)) 
- A_{\pm}(k) (B_{\pm}(p) - B_{\pm}(k) ) \bigr\}. 
\end{eqnarray}
It is expected that the gauge dependence is suppressed in the solution of Eqs.~(\ref{SD:A:BC}) and (\ref{SD:B:BC}) compared with the rainbow-ladder approximation.
At the limit $\mu \rightarrow 0$, the right-hand side of Eqs.~(\ref{SD:A:BC}) and (\ref{SD:B:BC}) have equivalent forms for each parities and independent on the sign for $B_\pm$.
Thus we find the solutions,
\begin{equation}
\left\{
\begin{array}{l}
A_{+}(p)=A_{-}(p) , \\
B_{+}(p)-m =\pm [B_{-}(p)+m] .
\end{array}
\right.
\label{sol:mu0}
\end{equation}
One of the solution, $B_{+}+m = B_{-}-m$, spontaneously breaks the chiral symmetry.

A simple relation, $Z_{1\pm} =Z_{2\pm}$, is known between the renormalization constant for the coupling constant, $Z_{1\pm}$, and the fermion field, $Z_{2\pm}$, from the consequence of the WT identity. The rainbow-ladder approximation satisfies the relation, $Z_{1\pm}=Z_{2\pm}$, in the Landau gauge at $\mu=0$. 
The relation, $Z_{1\pm}=Z_{2\pm}$, is maintained for a finite $\mu$ without the final term in the BC vertex function (\ref{BC}), see appendix. In this case we obtain a simpler form of the vertex function,
\begin{eqnarray}
\Gamma^{\mu}=\Gamma_{L+}^{(2) \mu}+\Gamma_{L-}^{(2) \mu},
\end{eqnarray}
where
\begin{eqnarray}
\Gamma_{L\pm}^{(2) \mu} = \frac{1}{2} \biggl\{ (A_{\pm}(p) + A_{\pm}(k))\gamma^{\mu}
+ (A_{\pm}(p) - A_{\pm}(k)) \frac{p^{\mu} + k^{\mu}}{p^2 - k^2}
(\slash{p} + \slash{k})  \biggr\} \chi_{\pm}.
\label{vertex:2}
\end{eqnarray}
In this vertex function it is possible to avoid a numerical instability introduced by a term proportional to $(B(p) - B(k))/(p^2-k^2)$ for $p\sim k$. The instability appears for a small topological mass, $\mu$. We discuss a concrete situation in the next section.

Substituting the vertex function (\ref{vertex:2}) in Eq.~(\ref{SD}) or eliminating terms proportional to $B(p) - B(k)$ in Eqs.~(\ref{SD:A:BC}) and (\ref{SD:B:BC}), we obtain the following simpler simultaneous equations:
\begin{eqnarray}
A_{\pm}(p) &=& 1- \frac{e^2}{8\pi^2} \int dk \frac{k}{A_{\pm}^2(k) k^2 + B_{\pm}^2(k)}
\nonumber\\
&&\times \biggl\{ F_{A\pm}(k,p,\mu) 
 + G_{A1\pm}(k,p,\mu) \ln \! \frac{(p-k)^2}{(p+k)^2} 
  + G_{A2\pm}(k,p,\mu) \ln \! \frac{(p-k)^2 + \mu^2}{(p+k)^2 + \mu^2} \nonumber\\
&& + G_{A3\pm}(k,p,\mu) 
\biggl( \ln \! \frac{(p-k)^2}{(p+k)^2} - \ln \! \frac{(p-k)^2 + \mu^2}{(p+k)^2 + \mu^2} \biggr)
\biggr\},
\label{SD:A:2}
\end{eqnarray}
with
\begin{eqnarray}
F_{A\pm}(k,p,\mu) &=& (1+\alpha) \frac{k}{p^2} A_{\pm}(k)(A_{\pm}(p) + A_{\pm}(k))\nonumber\\
&&+\frac{k}{p^2 (p^2 - k^2)} \bigl\{
 - (1 - \alpha ) (p^2+k^2) A_{\pm}(k)(A_{\pm}(p) - A_{\pm}(k)) \nonumber\\
&& \pm \mu B_{\pm}(k) (A_{\pm}(p) - A_{\pm}(k)) )
\bigr\},
\end{eqnarray}
\begin{eqnarray}
G_{A1\pm}(k,p,\mu) &=& \frac{\alpha}{2p^3} \biggl\{
 \frac{p^2+k^2}{2} A_{\pm}(k)(A_{\pm}(p) + A_{\pm}(k)) \nonumber\\
&& + \frac{p^2 - k^2}{2} A_{\pm}(k) (A_{\pm}(p) - A_{\pm}(k)) 
\bigr\},
\end{eqnarray}
\begin{eqnarray}
G_{A2\pm}(k,p,\mu) &=& \frac{\mu}{2p^3} \biggl\{
 \frac{\mu}{2} A_{\pm}(k)(A_{\pm}(p) + A_{\pm}(k)) 
  \pm B_{\pm}(k) (A_{\pm}(p) + A_{\pm}(k)) \biggr\} \nonumber\\
&+& \frac{\mu^2 + 2(p^2+k^2)}{2 p^3 (p^2-k^2)} \biggl\{
 - \frac{ p^2 + k^2 }{2} A_{\pm}(k)(A_{\pm}(p) - A_{\pm}(k)) \nonumber\\
&& \pm \frac{\mu}{2} B_{\pm}(k) (A_{\pm}(p) - A_{\pm}(k)) 
\biggr\},
\end{eqnarray}
\begin{eqnarray}
G_{A3\pm}(k,p,\mu) &=& \frac{p^2-k^2}{2\mu^{2}p^{3}} \biggl\{
 \frac{p^2-k^2}{2} A_{\pm}(k)(A_{\pm}(p) + A_{\pm}(k)) \nonumber\\
&& + \frac{ p^2 +k^2 }{2} A_{\pm}(k) (A_{\pm}(p) - A_{\pm}(k)) \nonumber\\
&& \pm \mu B_{\pm}(k) (A_{\pm}(p) + A_{\pm}(k)) 
\pm \frac{\mu}{2} B_{\pm}(k) (A_{\pm}(p) - A_{\pm}(k)) 
\biggr\}.
\end{eqnarray}
and
\begin{eqnarray}
B_{\pm}(p) &=& \pm m - \frac{e^2}{8\pi^2} \int dk \frac{k}{A_{\pm}^2(k) k^2 + B_{\pm}^2(k)}
\nonumber\\
&&\times \biggl\{ F_{B\pm}(k,p,\mu) 
 + G_{B1\pm}(k,p,\mu) \ln \! \frac{(p-k)^2}{(p+k)^2} 
  + G_{B2\pm}(k,p,\mu) \ln \! \frac{(p-k)^2 + \mu^2}{(p+k)^2 + \mu^2} \nonumber\\
&& + G_{B3\pm}(k,p,\mu) 
\biggl( \ln \! \frac{(p-k)^2}{(p+k)^2} 
       - \ln \! \frac{(p-k)^2 + \mu^2}{(p+k)^2 + \mu^2} \biggr)
\biggr\},
\label{SD:B:2}
\end{eqnarray}
with
\begin{eqnarray}
F_{B\pm}(k,p,\mu) &=& \frac{k}{p^2 - k^2} \bigl\{
\pm \mu A_{\pm}(k)(A_{\pm}(p) - A_{\pm}(k)) \nonumber\\
&& + 2(1-\alpha) B_{\pm}(k) (A_{\pm}(p) - A_{\pm}(k)) \bigr\}, 
\end{eqnarray}
\begin{eqnarray}
G_{B1\pm}(k,p,\mu) &=& \frac{\alpha}{2p} 
 B_{\pm}(k)(A_{\pm}(p) + A_{\pm}(k)),
 \end{eqnarray}
\begin{eqnarray}
G_{B2\pm}(k,p,\mu) &=& \frac{1}{p} \biggl\{
 \pm \frac{\mu}{2} A_{\pm}(k)(A_{\pm}(p) + A_{\pm}(k)) 
  + B_{\pm}(k) (A_{\pm}(p) + A_{\pm}(k)) \biggr\}\nonumber\\
&+& \frac{\mu^2 + 2(p^2+k^2)}{2p(p^2-k^2)} \biggl\{
 \pm \frac{\mu}{2} A_{\pm}(k)(A_{\pm}(p) - A_{\pm}(k)) \nonumber\\
&& + B_{\pm}(k) (A_{\pm}(p) - A_{\pm}(k)) \biggr\},
\end{eqnarray}
\begin{eqnarray}
G_{B3\pm}(k,p,\mu) &=& -\frac{p^2-k^2}{2\mu^2p} \bigl\{
 \pm \mu A_{\pm}(k)(A_{\pm}(p) + A_{\pm}(k)) \nonumber\\
&& \pm \frac{\mu}{2} A_{\pm}(k) (A_{\pm}(p) - A_{\pm}(k)) 
+ B_{\pm}(k) (A_{\pm}(p) - A_{\pm}(k)) \bigr\}.
\end{eqnarray}

In the following section, we numerically analyze the SD equation with the BC vertex function, $\Gamma_{\pm}^{BC \mu}$, a simpler one, $\Gamma_{\pm}^{(2) \mu}$, and the tree level one, $\gamma^{\mu}$ to derive information on the phase boundary and study the gauge dependence of the result. We are interested in the parity violating contribution for the fermion propagator from the CS term. Thus we eliminate the explicit parity breaking mass and set $m=0$.

\section{Numerical solutions}

We wish to study spontaneous breaking of the symmetry for the parity and the chiral transformation. It is obtained by observing the expectation values for the composite operators constructed by the fermion and the anti-fermion,
$\bar{\psi} \psi$ and $\bar{\psi} \tau \psi$.
If the operator, $\bar{\psi} \psi$, develops a non-vanishing expectation value, the chiral symmetry is spontaneously broken.
The operator, $\bar{\psi} \tau \psi$, is invariant under the chiral transformations (\ref{transformation:chiral}). The non-vanishing expectation value for $\bar{\psi} \tau \psi$ spontaneously breaks the parity conservation for $\mu=0$.
We evaluate the phase structure of the theory by observing the expectation values for $\bar{\psi} \psi$ and $\bar{\psi} \tau \psi$.

The fermion propagator is represented by the functions, $A_{\pm}(p)$ and $B_{\pm}(p)$ in Eq.~(\ref{pro:fermion}). These functions are calculated by solving the SD equation. The expectation values for $\bar{\psi} \psi$ and $\bar{\psi} \tau \psi$ are obtained by the solution of the SD equation,
\begin{eqnarray}
\langle \bar{\psi} \psi \rangle = -\frac{1}{\pi^2}\int dk
\left( \frac{k^2B_{+}(k) }{A_{+}^2(k)k^2 + B_{+}^2(k)} 
 + \frac{k^2B_{-}(k) }{A_{-}^2(k)k^2 + B_{-}^2(k)} \right),
\label{chiral}
\end{eqnarray}
and
\begin{eqnarray}
\langle \bar{\psi} \tau \psi \rangle = -\frac{1}{\pi^2}\int dk
\left( \frac{k^2B_{+}(k) }{A_{+}^2(k)k^2 + B_{+}^2(k)} 
 - \frac{k^2B_{-}(k) }{A_{-}^2(k)k^2 + B_{-}^2(k)} \right).
\label{parity}
\end{eqnarray}

There are some numerical methods available for solving the SD equation developed in the previous section. In this paper we employ a simple iteration which is used in Ref.~\cite{Fukazawa:1999aj}. We start with suitable trial function for the solution and iterating the calculation until stable solutions are obtained. It should be noted that the number of iteration to obtain the stable solutions increases near the critical point of the chiral symmetry breaking. 

The simplest choice for the trial functions is constants, independent of the momentum. For the rainbow-ladder approximation in the Landau gauge we start with the chiral symmetric and parity invariant trial functions,
\begin{eqnarray}
(a) & A_\pm (p)=1.0 , & \frac{B_\pm (p)}{M}=\pm 0.01 ,
\nonumber \\
(b) & A_\pm (p)=1.0 , & \frac{B_\pm (p)}{M}=\mp 0.01 ,
\label{start:iteration}
\end{eqnarray}
respectively. Here we normalize dimensionful quantities to the mass scale, $M$, defined by
\begin{eqnarray}
M\equiv e^2.
\end{eqnarray}
The solution in the Landau gauge is used as initial trial functions to calculate the SD equation at $\alpha=0.1$. The SD equation $\alpha=i+0.1$ is iteratively calculated starting from the solution at $\alpha=i$. We repeat the calculation and evaluate the gauge dependence of the solution.

To solve the SD equation for the vertex function, $\Gamma_{\pm}^{(2) \mu}$, we start with the trial functions (\ref{start:iteration}) 
and repeat the iteration until stable solutions are obtained. 
For the BC vertex function, $\Gamma_{\pm}^{(BC) \mu}$, we also start with the trial functions (\ref{start:iteration}). We repeat the iteration with eliminating the terms proportional to $B_\pm(k)(B_\pm(p)-B_\pm(k))$ for 10 times then introduce the BC vertex function, $\Gamma_{\pm}^{(BC) \mu}$. After sufficiently many iterations we obtain equivalent stable solutions for each trial functions (\ref{start:iteration}) if the chiral symmetry is restored.

To regularize the momentum integral, the UV cutoff scale, $\Lambda_{UV}$, is introduced. We also introduce the IR cutoff scale, $\Lambda_{IR}$, to eliminate the numerical error. These scales are fixed at $\Lambda_{IR}=10^{-4}M$ and $\Lambda_{UV}=10^{3}M$. The $k$-integration in Eqs.~(\ref{SD:A:BC}), (\ref{SD:B:BC}), (\ref{SD:A:2}) and (\ref{SD:B:2}) is performed by using the trapezoidal rule from $\Lambda_{IR}$ to $\Lambda_{UV}$. Since a non-trivial behavior is observed for $B_{\pm}(p)$ at low momentum, we want to sample low momentum interval more frequently and reduce the numerical cost. We define a variable, $x\equiv \ln p$, and divide the integration interval into 1000 subintervals of equal width with respect to the variable, $x$.

We have to be carefully treat a round-off error in the integrand. It can grow per iteration. A growing round-off error appears for the following three cases,
\begin{eqnarray}
& (i) & p/k < \varepsilon (=0.1), \nonumber \\
& (j) & k/p < \varepsilon (=0.1), \nonumber \\
& (k) & k/\mu < \varepsilon (=0.1)\, \, \mbox{and}\, \, k/p \sim \mbox{O}(1). \nonumber
\end{eqnarray}
We expand the integrand in terms of the small variables, $p/k$, $k/p$ and $k/\mu$ and use the approximate equation up to 6th order of the small variables in each cases. In the case $3$ the terms proportional to $B_{\pm}(k)(B_{\pm}(p)-B_{\pm}(k))$ generate much larger growing round-off error in Eq.~(\ref{SD:A:BC}). To obtain a stable solution we assume that the small $B_{\pm}(p)-B_{\pm}(k)$ has no major contribution and ignore it for $|B_{\pm}(p)-B_{\pm}(k)|< 0.01M$.

\begin{figure}[htb]
 \begin{minipage}{0.5\hsize}
	\centering\includegraphics[width=2.5in]{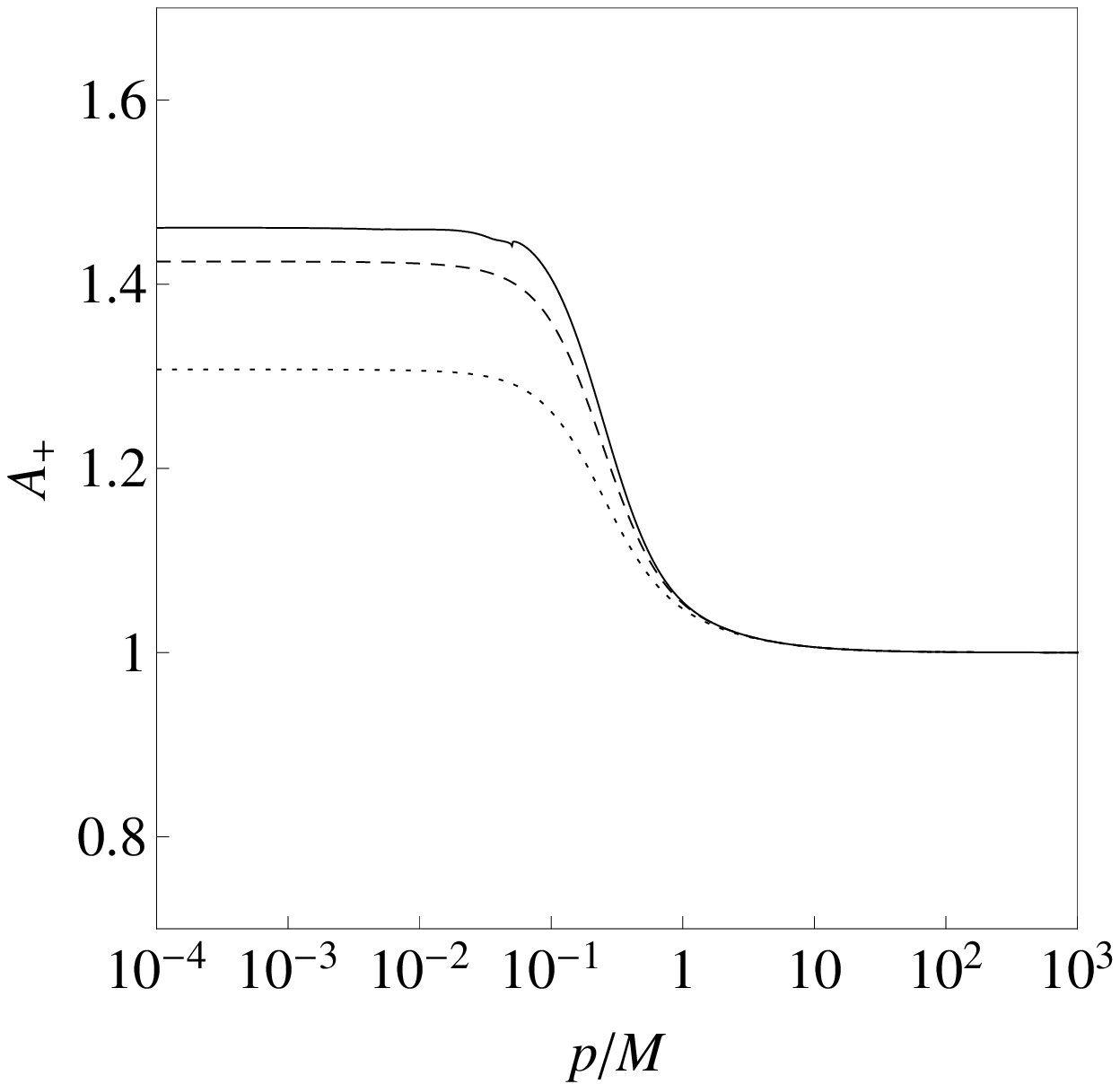}
 \end{minipage}
 \begin{minipage}{0.5\hsize}
	\centering\includegraphics[width=2.5in]{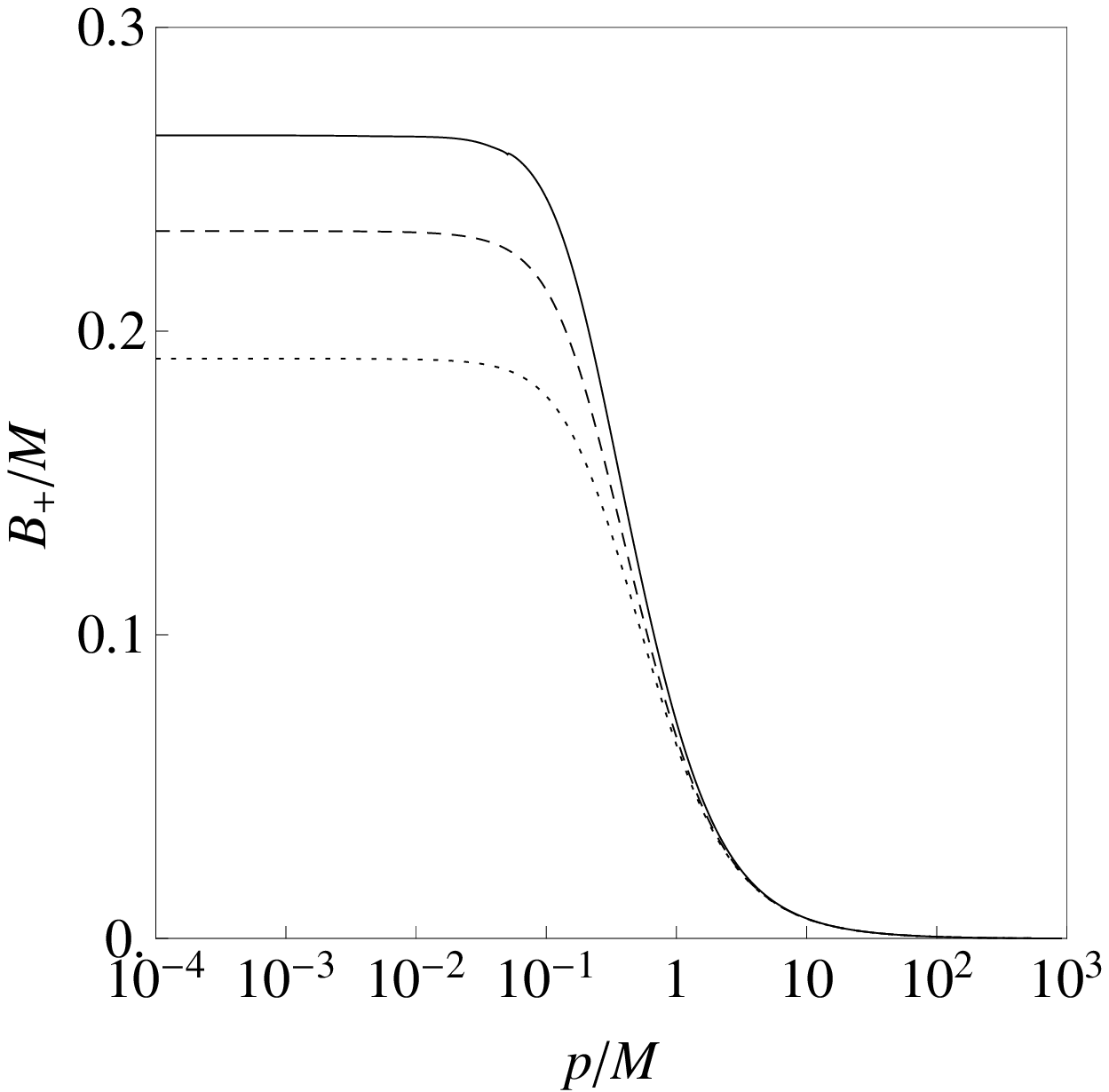}
\end{minipage}
	\caption{Typical shapes for the solutions, $A_{+}(p)$ and $B_{+}(p)/M$, for $\alpha=1$ and $\mu/M=0.5$. 
    The dotted, dashed and solid lines show the solutions in the rainbow-ladder approximation, the vertex function, $\Gamma_\pm^{(2)\mu}$, and the BC vertex function, $\Gamma_\pm^{BC\mu}$, respectively. }
	\label{Fig_AB}
\end{figure}

In Fig.~\ref{Fig_AB} we take the Feynman gauge $\alpha=1$ at $\mu/M=0.5$ and illustrate the functions, $A_{+}(p)$ and $B_{+}(p)/M$, as functions of the momentum $p$ normalized by $M$. The functions, $A_{+}(p)$ and $B_{+}(p)/M$, are seems to be smooth functions in terms of the momentum $p$.  A small crack is observed in the solid line for $A_{+}(p)$ from $p/M=10^{-2}$ to $p/M=10^{-1}$. It comes from our algorithm to reduce the round-off error. If we take the Landau gauge, the solutions in the rainbow-ladder approximation and the vertex function, $\Gamma_\pm^{(2)\mu}$ approaches to the one obtained with the BC vertex function. Tree lines almost overlap each other at $\alpha=0$. The chiral symmetry is restored at$\mu/M=0.5$. Thus we find only the solution where the functions, $A_{-}(p)$ and $B_{-}(p)/M$, coincide with $A_{+}(p)$ and $-B_{+}(p)/M$.

\begin{figure}[htb]
 \begin{minipage}{0.5\hsize}
	\centering\includegraphics[width=2.5in]{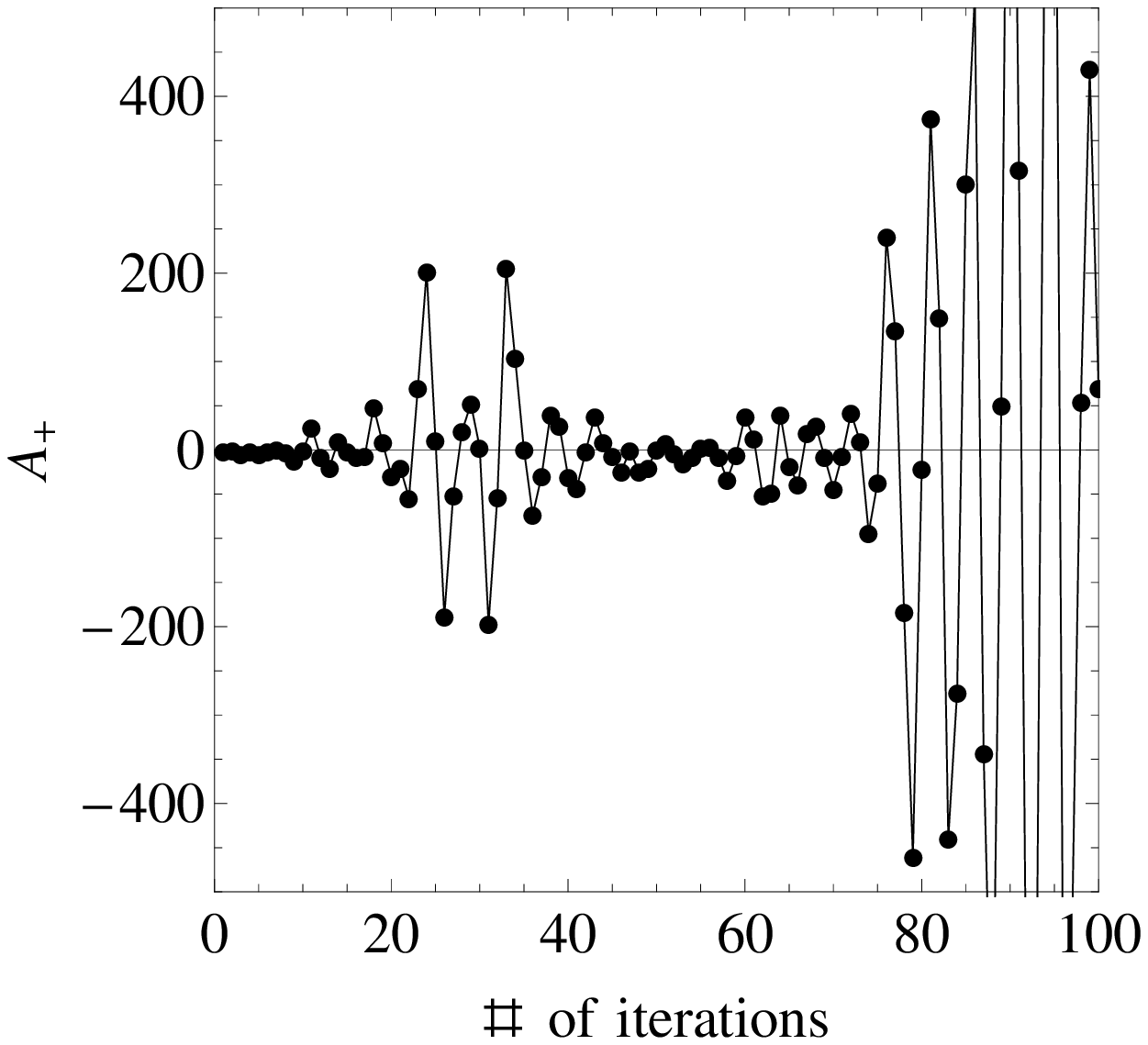}
 \end{minipage}
 \begin{minipage}{0.5\hsize}
	\centering\includegraphics[width=2.5in]{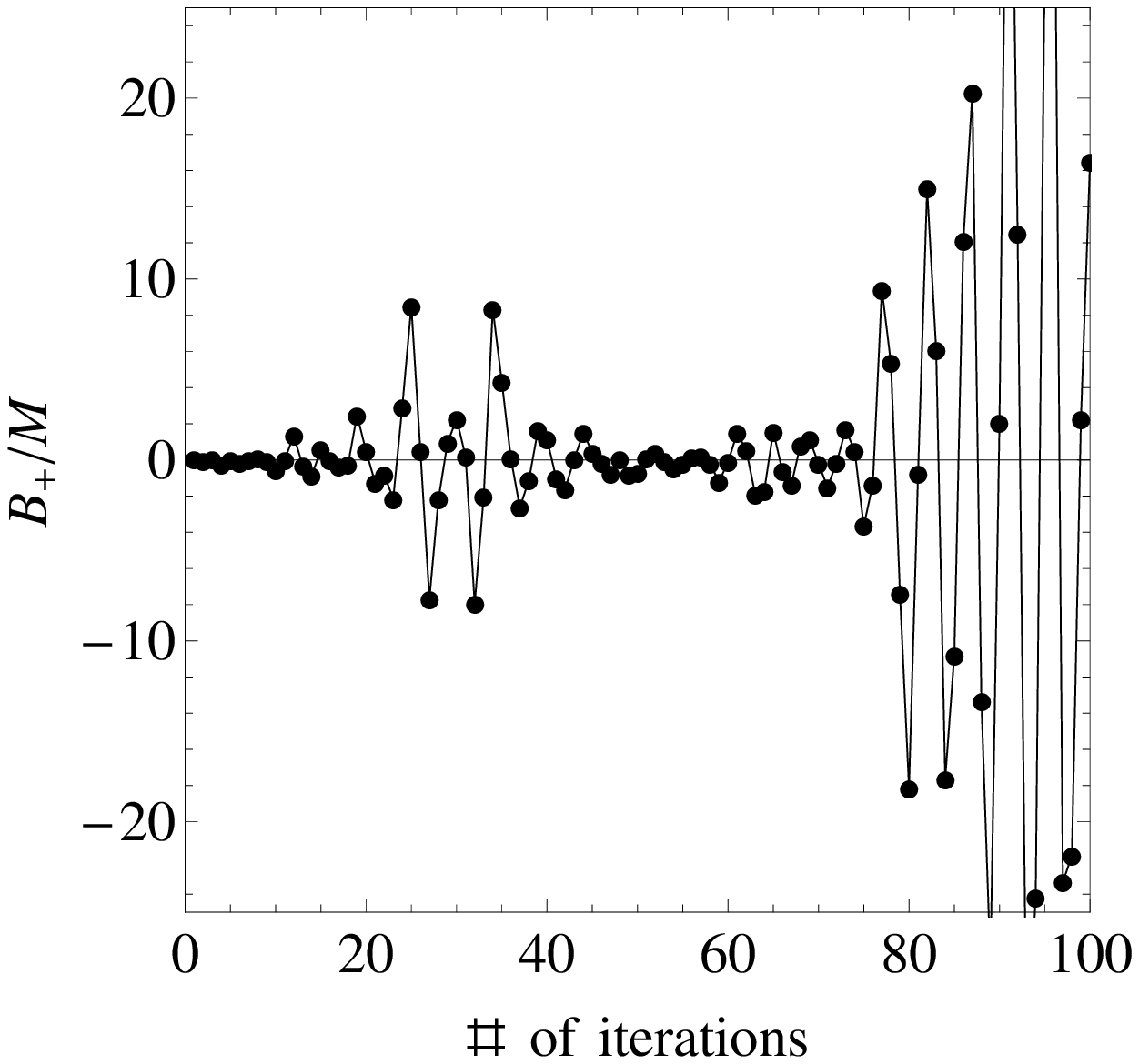}
 \end{minipage}
	\caption{Typical behavior for the solutions, $A_{+}(p)$ and $B_{+}(p)/M$, at $\alpha=0$. $\mu/M=0.02$ and $p/M=10^{-3}$ with the BC vertex function. }
	\label{Fig_Iteration}
\end{figure}

For a small $\mu$ we can not find any stable solution in the case with the BC vertex function. In Fig.~\ref{Fig_Iteration} the function, $A_{+}(p)$ and $B_{+}(p)/M$ are presented as a function of the number of iterations for $p/M=10^{-3}$. An instability for $A_+(p)$ generates a fluctuation for $B_+(p)$. In Fig.\ \ref{Fig_Iteration} a large fluctuation is observed for $B_+(p)$ one step later than $A_+(p)$. Observed fluctuations are much larger than typical values for $A_{+}(p)$ and $B_{+}(p)/M$ and do not converge after many iterations. It should be noted that the convergence of the iteration becomes slower near the critical value for $\mu$ even in the rainbow-ladder approximation. In this case more iterations are necessary to obtain stable solutions.

\section{Phase Structure}
The phase structure of the theory is found by evaluating the expectation values for $\bar{\psi} \psi$ and $\bar{\psi} \tau \psi$.
Since the these expectation values are physical observables, they have to be invariant under the gauge transformation.
Here we approximately solve the SD equation and insert the solution into Eqs. (\ref{chiral}) and (\ref{parity}).
The approximation introduces the gauge dependence in the result. It should be noted that the functions $A_{\pm}(p)$ and $B_{\pm}(p)$ are not necessary to be gauge invariant.

\begin{figure}[htb]
	\centering\includegraphics[width=4in]{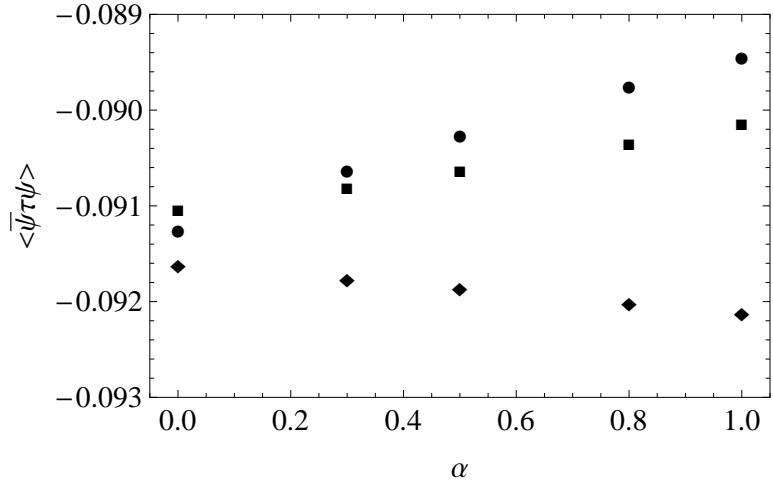}
	\caption{The behavior of the order parameter, $\langle \bar{\psi} \tau \psi \rangle$ as a function of the gauge parameter, $\alpha$ at $\mu/M=0.5$. The circle, triangle and square show the results for the vertex functions, $\gamma^\mu$, $\Gamma_{\pm}^{(2) \mu}$ and $\Gamma_{\pm}^{BC \mu}$, respectively.}
	\label{Fig_alp_dep1}
\end{figure}

For a small $\mu$ the numerical iteration does not always converge by using the vertex functions, $\Gamma_{\pm}^{BC \mu}$ and $\Gamma_{\pm}^{(2) \mu}$. First we calculate the expectation values, $\langle \bar{\psi} \psi \rangle$ and $\langle \bar{\psi} \tau \psi \rangle$, at $\mu/M=0.5$ to check the gauge dependence of the result for each vertex functions. The numerical solution for the order parameter, $\langle \bar{\psi} \psi \rangle$ vanishes at $\mu/M=0.5$. The chiral symmetry is not broken spontaneously. The CS term explicitly breaks the parity invariance and induces non-vanishing values for the order parameter, $\langle \bar{\psi} \tau \psi \rangle$. The behavior of the order parameter is plotted for each vertex functions in Fig. \ref{Fig_alp_dep1}.

The vertex functions, $\Gamma_{\pm}^{BC \mu}$ and $\Gamma_{\pm}^{(2) \mu}$ show smaller gauge dependences than the tree level one. In Fig. \ref{Fig_alp_dep1} almost proportional relation is observed between the order and the gauge parameters. Assuming the relation
\begin{eqnarray}
\langle \bar{\psi}\tau\psi \rangle -  \langle \bar{\psi}\tau\psi \rangle|_{\alpha =0 } 
= a \alpha + O(\alpha^2),
\end{eqnarray}
we estimate the slope $a$ by the least-squares method.
\begin{table}
\caption{The gauge dependence for the parity violating order parameter at $\mu/M=0.5$.}
\label{slope}
\begin{center}
  \begin{tabular}{c|c}
 Vertex function & $a$ \\
\hline
 $\gamma^\mu$ & $(1.79 \pm 0.06) \times 10^{-3}$ \\
 $\Gamma_{\pm}^{(2) \mu}$ & $(0.91 \pm 0.03) \times 10^{-3}$ \\
 $\Gamma_{\pm}^{BC \mu}$ & $-(0.495 \pm 0.005) \times 10^{-3}$
  \end{tabular}
\end{center}
\end{table}
From Tab. \ref{slope} we clearly find the smallest gauge dependence for the BC vertex function.

As is shown in Fig. \ref{Fig_alp_dep1}, the difference of each vertex functions decreases for small $\alpha$. The contribution from the vertex correction almost disappear near the Landau gauge, $\alpha=0$. Below we fix the gauge parameter as $\alpha=0$ and continue the analysis for a smaller topological mass, $\mu$.

\begin{figure}[htb]
	\centering\includegraphics[width=4in]{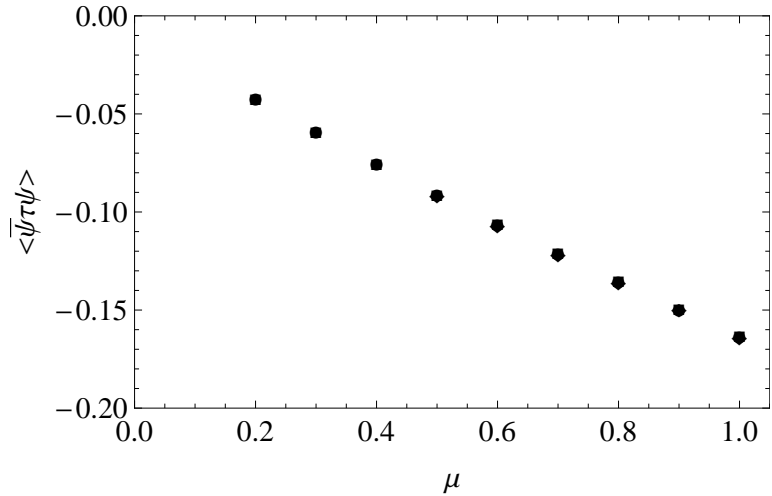}
	\caption{The behavior of the order parameter, $\langle \bar{\psi} \tau \psi \rangle$ in the Landau gauge. The circle, triangle and square show the results for the vertex functions, $\gamma^\mu$, $\Gamma_{\pm}^{(2) \mu}$ and $\Gamma_{\pm}^{BC \mu}$, respectively.}
	\label{Fig_mu_dep_PB2}
\end{figure}

The order parameter, $\langle \bar{\psi} \tau \psi \rangle$ are plotted for $0.1 \leq \mu/M \leq 1.0$ in Fig. \ref{Fig_mu_dep_PB2}. 
A larger absolute value is observed for $\langle \bar{\psi} \tau \psi \rangle$ as the topological mass, $\mu$, increases.
It is a direct consequence from the explicit parity violation of the CS term.
In the Landau gauge each vertex functions generate nearly equivalent expectation values for $0.1 \leq \mu/M \leq 1.0$.

\begin{figure}[t]
	\centering\includegraphics[width=4in]{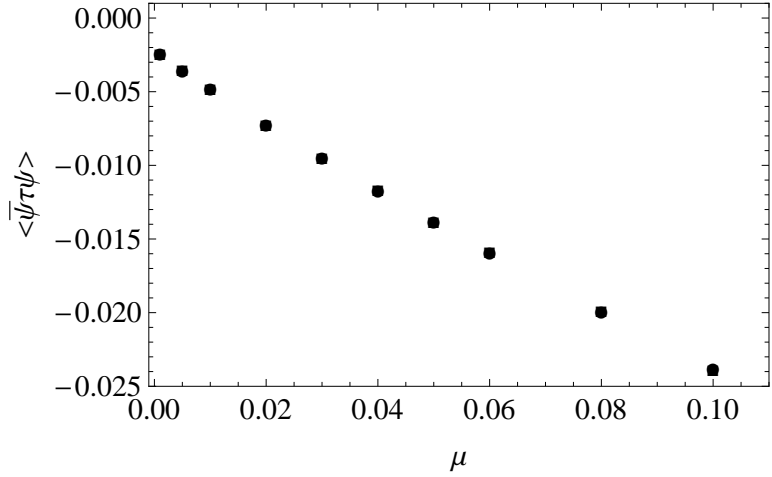}
	\caption{The behavior of the order parameter, $\langle \bar{\psi} \tau \psi \rangle$ in the Landau gauge. The circle and triangle show the results for the vertex functions, $\gamma^\mu$ and $\Gamma_{\pm}^{(2) \mu}$, respectively.}
	\label{Fig_mu_dep_PB1}
\end{figure}

Starting from different initial values for $A_{\pm}(p)$ and $B_{\pm}(p)$, we obtain two independent solutions of the SD equation for a smaller $\mu$. It should be noted that two solutions are also obtained by introducing a finite fermion mass in Ref. \cite{Zhu:2013zna}.
In our case the one solution preserves the chiral symmetry, $\langle \bar{\psi} \psi \rangle =0$. In this case the expectation value for $\langle \bar{\psi} \tau \psi \rangle$ develops a non-vanishing value at the limit $\mu \rightarrow 0$, as is shown in Fig.~\ref{Fig_mu_dep_PB1}. In this figure we plot the results of the vertex functions, $\gamma^\mu$ and $\Gamma_{\pm}^{(2) \mu}$ in the Landau gauge.  In the case of the BC vertex function the iteration does not converge for a smaller $\mu$.

\begin{figure}[htb]
	\centering\includegraphics[width=4in]{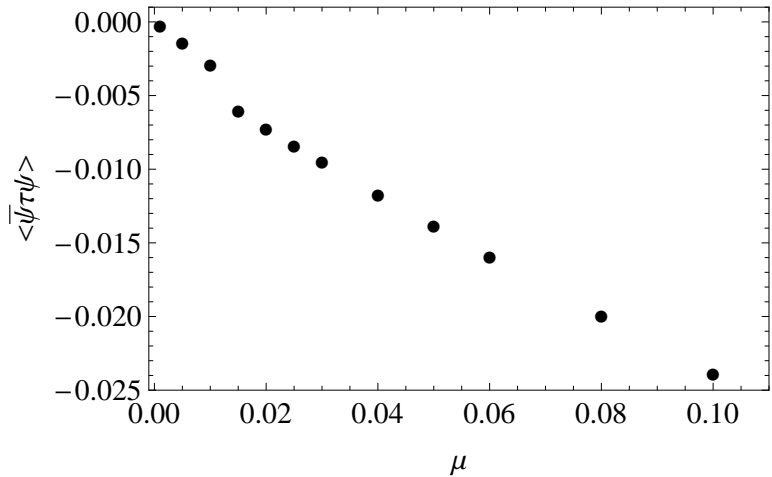}
	\caption{The behavior of the order parameter, $\langle \bar{\psi} \tau \psi \rangle$ for the rainbow-ladder approximation in the Landau gauge.}
	\label{Fig_v0_CB_P}
\end{figure}

\begin{figure}[htb]
	\centering\includegraphics[width=4in]{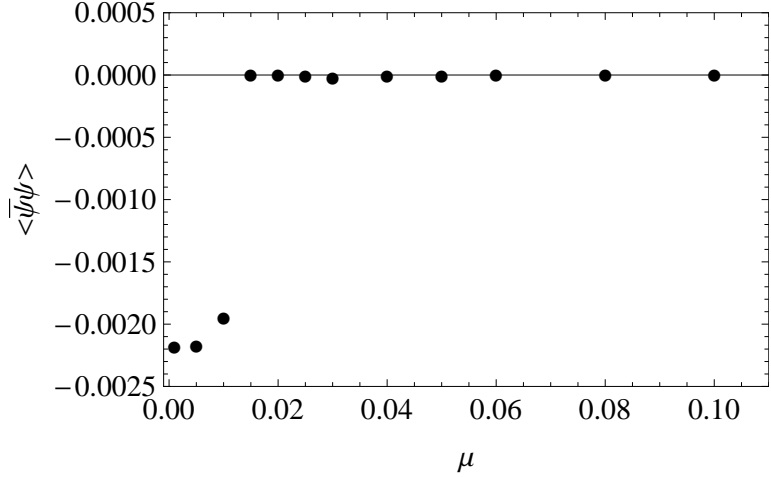}
	\caption{The behavior of the order parameter, $\langle \bar{\psi} \psi \rangle$ for the rainbow-ladder approximation in the Landau gauge.}
	\label{Fig_v0_CB_P2}
\end{figure}

In the other solution the expectation value for $\langle \bar{\psi} \tau \psi \rangle$ vanishes at the limit $\mu \rightarrow 0$. The CS term disappears at the limit. We set the parity violating fermion mass as $m=0$. Since there is no explicit parity breaking, this solution seems to be consistent with our physical set up. The solution exhibits the same qualitative features as those obtained in Ref.~\cite{Madrigal:2011wr}. 

The figures \ref{Fig_v0_CB_P} and \ref{Fig_v0_CB_P2} show the behavior of the order parameters, $\langle \bar{\psi} \tau \psi \rangle$ and $\langle \bar{\psi} \psi \rangle$. As is clearly seen in Fig.~\ref{Fig_v0_CB_P2}, the chiral order parameter has a non-vanishing value for a small topological mass. The broken chiral symmetry is restored at $\mu = \mu_C \sim 0.01M$. The phase transition from the broken phase to the symmetric phase seems to be of the first order \cite{Kondo:1994bt}. We also observe a gap like structure at $\mu = \mu_C$ for $\langle \bar{\psi} \tau \psi \rangle$ in Fig.~\ref{Fig_v0_CB_P}.

We note that a stable solution can be found only for the rainbow-ladder approximation near the critical topological mass, $\mu\sim\mu_c$. There is numerical difficulty to find a solution in the vertex functions $\Gamma_{\pm}^{(2) \mu}$ and $\Gamma_{\pm}^{BC \mu}$.  Thus we plot the result for the rainbow-ladder approximation in the Landau gauge in Figs.~\ref{Fig_v0_CB_P} and \ref{Fig_v0_CB_P2}. However, the contribution from the vertex function is small enough in the Landau gauge for a larger topological mass and the gauge dependence of the result is suppressed in the BC vertex function. 

The solutions with the vertex functions $\Gamma_{\pm}^{(2) \mu}$ satisfies Eq.~(\ref{sol:mu0}) at $\mu=0$. The expectation value, $\langle \bar{\psi} \psi \rangle$,  for the solution, $B_{+} = B_{-}$, coincides with the value, $\langle \bar{\psi} \tau \psi \rangle$ for the solution, $B_{+} = - B_{-}$, which is shown in Fig.~\ref{Fig_mu_dep_PB1}. It is nearly equivalent to the solution for the rainbow-ladder approximation in the Landau gauge at $\mu=0$. We consider that the result in the Landau gauge is consistent with the gauge invariance.

\section{Conclusion}

We have investigated the phase structure for the four component fermion in an Abelian gauge theory with CS term in 1+2 dimensions. The SD equation was solved numerically with the vertex functions that are consistent with the WT identity. We evaluated two types of the expectation values and found the critical topological mass, $\mu_c$. The gauge dependence of the result was also clarified.

The BC vertex function reduces to about half the gauge dependence for the solution of the SD equation compared with the rainbow-ladder approximation. The contribution from the vertex correction suppressed near the Landau gauge, $\alpha=0$. The differences of the each vertex functions are small enough in the Landau gauge. Thus we conclude that the rainbow-ladder approximation is valid in the Landau gauge even for a finite $\mu$ case where the WT identity is not satisfied.

The CS term explicitly breaks the parity invariance. It generates the expectation value of the parity operator, $\langle \bar{\psi} \tau \psi \rangle$. The absolute value of $\langle \bar{\psi} \tau \psi \rangle$ monotonically increases as a function of the topological mass, $\mu$. For a small $\mu$ the SD equation has two independent solutions. The one shows the chiral symmetry breaking for $\mu < \mu_c$. A gap like structure is observed for $\langle \bar{\psi} \tau \psi \rangle$ at $\mu < \mu_c$. In the other solution the chiral symmetry is not broken. There is no gap like structure for $\langle \bar{\psi} \tau \psi \rangle$.

Although this work is restricted to the analysis of the expectation values for fermion bi-linear operators, we expect that the rainbow-ladder approximation in the Landau gauge is valid in the general case.
We are interested in applying the procedure to the gauge theory at finite temperature where the Ward identity, $Z_1=Z_2$, is broken in the rainbow-ladder approximation with a constant gauge parameter.

In our analysis we suppose a simple form for the vertex function and impose $\Gamma_{T}^{\mu} = 0$. Since the transverse part of the vertex function is generated at one loop level \cite{ConchaSanchez:2013cp}, we should consider a general form for the vertex function to apply our result to some specific systems.
We will continue the work further and hope to report on these problems.

\section*{Acknowledgment}

The authors thank Wataru Sakamoto for a preliminary contribution at the early stage of this work. 
TI is supported by JSPS KAKENHI Grant Number 26400250. 


%


\appendix
\section{Ward identity}

We introduce the renormalization constants, $Z_{1\pm}$, $Z_{2\pm}$, and redefine the fields in the following way,
\begin{eqnarray}
e\bar{\psi}\gamma^{\rho}\psi A_{\rho}
&=&Z_{1+}e_r\bar{\psi_r}\gamma^{\rho}\chi_{+}\psi_r A_{r\rho}
+Z_{1-}e_r\bar{\psi_r}\gamma^{\rho}\chi_{-}\psi_r A_{r\rho},
\nonumber \\
\psi &=& Z_{2+}\chi_+\psi_r + Z_{2-}\chi_-\psi_r.
\end{eqnarray}

The renormalization constants are fixed by the renormalization condition.
We adopt the renormalization condition for the fermion propagator,
\begin{eqnarray}
\left. S(p)\right|_{p^2=0} = \frac{iZ_{2+}}{\slash{p}  - m_{r+} }\chi_{+}
 + \frac{iZ_{2-}}{\slash{p}  - m_{r-} }\chi_{-},
\label{pro:fermion:ren:cond}
\end{eqnarray}
where $m_{r\pm}$ denotes the renormalized mass. From Eqs.~(\ref{pro:fermion}) and (\ref{pro:fermion:ren:cond}) we find
\begin{eqnarray}
Z_{2\pm} = A^{-1}_{\pm}(p^2=0).
\label{Z2}
\end{eqnarray}
Imposing the renormalization condition
\begin{eqnarray}
\left. \Gamma^\mu(p,k)\right|_{p=k, p^2=k^2=0} = 
(Z_{1+}^{-1} \chi_+ + Z_{1-}^{-1} \chi_-) \gamma^\mu,
\end{eqnarray}
and substituting Eq. (\ref{vertex:2}) into this equation, we obtain
\begin{eqnarray}
Z_{1\pm} = A^{-1}_{\pm}(p^2=0),
\end{eqnarray}
with
\begin{eqnarray}
\left. \frac{\partial A_{\pm}(p^2)}{\partial p^2} p^{\mu}p^{\nu}\right|_{p^2=0}=0.
\label{rg:con}
\end{eqnarray}
In Euclidean space Eq.(\ref{rg:con}) reduces to
\begin{eqnarray}
\left. \frac{\partial A_{\pm}(p^2)}{\partial \mbox{ln} p^2} \right|_{p^2=0}=0,
\label{rg:con2}
\end{eqnarray}
From Fig.~\ref{Fig_AB} we observe that $A_{\pm}(p^2)$ is almost flat with respect to ln$p^2$ at the limit $p^2 =0$. Therefore the Ward identity, $Z_{1\pm}=Z_{2\pm}$, is satisfied for the vertex functions $\Gamma_{\pm}^{(2) \mu}(p,k)$.

We set $\Gamma^\mu(p,k)=\gamma^\mu$ in the rainbow-ladder approximation. It fixes the renormalization constants, $Z_{1\pm} = 1$. On the other hand the renormalization constants, $Z_{2\pm}$, are determined by Eq.(\ref{Z2}). Since the CS term shift the value, $A(0)$, from a unity \cite{Madrigal:2011wr}, the Ward identity can not be preserved for $\mu\neq 0$.

\end{document}